# Gate-controlled anisotropy in Aharonov-Casher spin interference: signatures of Dresselhaus spin-orbit inversion and spin phases

Fumiya Nagasawa[1], Andres A. Reynoso[2,3], José Pablo Baltanás[4], Diego Frustaglia[4,5], Henri Saarikoski[6], and Junsaku Nitta[1,7,8]

[1]Department of Materials Science, Tohoku University, Sendai, 980-8579, Japan
[2]Instituto Balseiro and Centro Atómico Bariloche, Comisión Nacional de Energía Atómica, 8400 Bariloche, Argentina
[3]Consejo Nacional de Investigaciones Científicas y Técnicas (CONICET), Argentina
[4]Departamento de Física Aplicada II, Universidad de Sevilla, E-41012, Sevilla, Spain
[5]Freiburg Institute for Advanced Studies (FRIAS), Albert-Ludwigs Universität Freiburg, D-79104 Freiburg, Germany
[6]RIKEN Center for Emergent Material Science (CEMS), Wako Saitama 351-0198, Japan
[7]Center for Spintronics Research Network, Tohoku University, Sendai, 980-8577, Japan.
[8]Organization for Advanced Studies, Center for Science and Innovation in Spintronics (Core Research Cluster), Tohoku University, Sendai, 980-8577, Japan

The coexistence of Rashba and Dresselhaus spin-orbit interactions (SOIs) in semiconductor quantum wells leads to an anisotropic effective field coupled to carriers' spins. We demonstrate a gate-controlled anisotropy in Aharonov-Casher (AC) spin interferometry experiments with InGaAs mesoscopic rings by using an in-plane magnetic field as a probe. Supported by a perturbation-theory approach, we find that the Rashba SOI strength controls the AC resistance anisotropy via spin dynamic and geometric phases and establish ways to manipulate them by employing electric and magnetic tunings. Moreover, assisted by two-dimensional numerical simulations, we identify a remarkable anisotropy inversion in our experiments attributed to a sign change in the renormalized linear Dresselhaus SOI controlled by electrical means, which would open a door to new possibilities for spin manipulation.

**I. INTRODUCTION**

Spintronics and spin-based quantum computing rely on the precise manipulation of spin orientations and related spin phases. Electron spins may couple directly to a magnetic field (Zeeman interaction) as well as to an electric field via spin-orbit interaction (SOI), resulting in a momentum-dependent effective magnetic field acting on itinerant spins. In particular, the electric-field-controllable Rashba SOI [1, 2, 3] is a prominent resource for *spin-orbitronics* [4], i.e., for the generation [5, 6, 7], manipulation [8, 9], and detection [10, 11] of spins by electrical means only. The direction of the effective Rashba field is perpendicular to the momentum of the spin carriers, but its strength is isotropic. In III-V compound semiconductors, the Dresselhaus SOI [12] induced by bulk inversion asymmetry also plays an important role in spin dynamics [13]. The direction of the effective Dresselhaus field has a different symmetry from Rashba's one. Therefore, the combination of Rashba and Dresselhaus SOIs gives rise to an anisotropic, momentum-dependent field.

A spin interferometer is an invaluable tool to probe the spin-phase information carried by electrons via the Aharonov-Casher (AC) effect [14, 15], the electromagnetic dual of the Aharonov-Bohm (AB) effect [16, 17]. The role played by Rashba and Zeeman fields on spin phases has been widely investigated in this context [18]-[22]. In contrast, the effects of introducing Dresselhaus SOI on spin phases are not yet well understood. Since hybrid-field engineering is a prerequisite to attain spin manipulation at the nanoscale, the electric control of the Dresselhaus SOI *strength and sign* appears as a challenging goal that would supply us with new tools for efficient spin control.

In this paper, we use Aharonov-Casher (AC) spin interferometry to extract information about the spin-orbit fields and related spin phases. We study the anisotropic response of AC resistance measurements in an array of InGaAs-based mesoscopic rings subject to in-plane magnetic fields oriented along different directions. The experiment shows that the sign of the AC resistance anisotropy changes as a function of the Rashba SOI strength. Perturbation-theory calculations indicate that the AC resistance anisotropy is modulated by the Rashba SOI strength via spin dynamic and geometric phases as well as by the direction of the in-plane Zeeman field [Section II & Appendix A2]. In addition, we find that the reported data are to a great extent reproduced by numerical results performed at constant Dresselhaus SOI strength [23]. There is, however, a remarkable discrepancy: the experiment reveals an extra sign inversion in the anisotropy which is not reproduced by the numerical calculations. This is consistently explained by a sign change of the renormalized linear Dresselhaus SOI emerging from strain effects in the working material, which is controlled electrically. Our results provide crucial information about the SO fields and show how different spin-phase contributions can be manipulated, demonstrating a potential for applications in spintronics and spin-based quantum technologies.

In the following Sec. II, the concept of spin dynamic and geometric phases in magnetic textures is introduced. In Sec. III, we show the anisotropic response of these phases when perturbed by additional Dresselhaus and Zeeman terms. The analytical details on the perturbation theory is described in Appendix A. In Sec. IV, we describe the gate-controlled anisotropy in Aharonov-Casher (AC) spin interferometry experiments with InGaAs mesoscopic rings by using an in-plane magnetic field as a probe. In Sec. V, we discuss a sign change of the renormalized linear Dresselhaus SOI. Section VI summarizes the paper.

**II. SPIN DYNAMICS IN MAGNETIC TEXTURES**

A magnetic texture is a magnetic field, either of real or effective (e.g., spin-orbit) origin, with non-uniform orientation. The spin dynamics of a carrier travelling through a magnetic texture is determined by the ratio of two characteristic frequencies: the Lamor frequency of spin precession around the local magnetic field, $\omega_s$, and an orbital frequency accounting for the change of direction of the magnetic field from the point of view of the spin carrier, $\omega_c$ [24]. The spin dynamics is said to be adiabatic if the carrier's spin can stay (anti)align with the local magnetic field all across the magnetic texture. This corresponds to the regime where the spin precession frequency is much larger than the orbital frequency, $\omega_s \gg \omega_c$. In the adiabatic limit, spin states have been shown to acquire phase contributions of geometric nature in addition to the usual dynamic quantum phases [25]. These geometric (or Berry) phases are identified with the solid angle $\Omega_s$ subtended by spins after a round trip in the Bloch sphere (spin texture). However, the adiabatic limit is difficult to achieve in usual experimental setups, where both frequencies tend to be of comparable magnitude and the spin dynamics is non-adiabatic. Still, geometric phases can be generalized to non-adiabatic situations with identical interpretation in terms of spin solid angles [26] even when the non-adiabatic spin texture does not coincide with the magnetic texture. Complementary dynamic spin phases are identified with the projection of the spin texture on the magnetic texture. See Fig. 1 for an illustration in the case of AC rings from the point of view of the spin carrier's rest frame.

**III. NONDEGENERATE PERTURBATION THEORY OF ANISOTROPIC SPIN INTERFERENCE**

The coexistence of Rashba and Dresselhaus SOIs leads to an anisotropic effective magnetic field $B_{SOI}(\varphi_0)$ with two-fold symmetry given by

$$g\mu_B B_{SOI}(\varphi_0) = 2k_F\sqrt{(\alpha^2 + \beta^2) - 2\alpha\beta \sin 2\varphi_0} \,, \qquad (1)$$

with g the gyromagnetic factor, $\mu_B$ the Bohr magneton, $\alpha$ the strength of Rashba SOI, $\beta = \gamma\langle k_z^2 \rangle$ the linear Dresselhaus SOI strength, $\gamma$ the bulk Dresselhaus SOI parameter, and $\langle k_z^2 \rangle$ resulting from the confinement of the wave vector in a two-dimensional (2D) quantum well (QW). Here, $\varphi_0$ is the direction of the electronic momentum with respect to the [100] direction, so that $k_x = k_F \cos\varphi_0$ and $k_y = k_F \sin\varphi_0$. The anisotropic effective field has maxima and minima at either $\varphi_0 = \pi/4$ or $\varphi_0 = 3\pi/4$ depending on the sign of $\alpha$ and $\beta$. This anisotropy can be tested by introducing an external in-plane Zeeman field as a probe. By treating the Dresselhaus SOI and the in-plane Zeeman field $B_{//}$ as perturbations to a large Rashba SOI, the time reversal AC conductance of ring-shaped spin interferometers [8] is given by [Appendix A2]

$$G_{AAS} = \frac{e^2}{h}\left[1 - \cos\left\{2\pi\left(\sqrt{1+\tilde{\alpha}^2} - 1 + \phi_B + \phi_D + \phi_A\right)\right\}\right], \quad (2)$$

with $\tilde{\alpha} = 2\alpha m^* r/\hbar^2$, $m^*$ the effective mass, $\hbar = h/2\pi$ the reduced Planck constant, and $r$ the radius of the ring. The corresponding phases read

$$\phi_B = \left(\frac{\omega_B}{\omega_0}\right)^2 \frac{1}{4(k_F r)^2 |\tilde{\alpha}|}, \quad \phi_D = \left(\frac{\omega_D}{\omega_0}\right)^2 \frac{1}{4|\tilde{\alpha}|},$$

and

$$\phi_A = -\left(\frac{\omega_B}{\omega_0}\right)^2 \left(\frac{\omega_D}{\omega_0}\right) \frac{\text{sign}[\tilde{\alpha}]}{8(k_F r)^2 \tilde{\alpha}^2} \sin 2\varphi,$$

with

$$\omega_0 = \hbar/m^* r^2, \quad \omega_B = 2\mu_B B_{//}/\hbar, \quad \omega_D = 2\beta/\hbar r, \quad (3)$$

where $\phi_B$ is the 2nd-order-perturbation Zeeman phase shift reported in [22], which was demonstrated to be of purely geometric origin. In contrast, the corresponding 2nd-order-perturbation Dresselhaus phase shift $\phi_D$ shows a hybrid geometric/dynamic origin [Appendix A2]. This is also the case for the 3rd-order-perturbation spin phase $\phi_A$ [Appendix A2], which is responsible for the anisotropic response of the conductance to the in-plane Zeeman field's direction $\varphi$, defined with respect to the [100] direction. Moreover, its linear dependence on $\omega_D$ shows that the anisotropic $\phi_A$ is sensitive to a sign inversion of the Dresselhaus SOI.

The AC conductance anisotropy can be studied by defining $A_G = G(\pi/4) - G(3\pi/4)$, the conductance difference for in-plane Zeeman fields $B_{//}$ oriented along different symmetry axes. The resulting expression in this approximation is

$$A_G = 2\sin(2\pi\tilde{\phi}_A)\sin\left[2\pi\left(\sqrt{1+\tilde{\alpha}^2} - 1 + \phi_B + \phi_D\right)\right]$$

with
$$\tilde{\phi}_A = -\left(\frac{\omega_B}{\omega_0}\right)^2 \left(\frac{\omega_D}{\omega_0}\right) \frac{\text{sign}[\tilde{\alpha}]}{8(k_F r)^2 \tilde{\alpha}^2} . \qquad (4)$$

The anisotropy $A_G$ oscillates as a function of $\alpha$, $\beta$, and $B_{//}$ as the corresponding phases increase. However, within this perturbative regime, only the dominating Rashba AC phase $\sqrt{1+\tilde{\alpha}^2}-1$ is expected to induce a sign inversion of $A_G$ as the Rashba SOI strength changes (a sign inversion due to the Zeeman field beyond the perturbative approach was confirmed by numerical analysis in [23]). Moreover, $\tilde{\phi}_A$ shows that an additional sign inversion is expected in $A_G$ in case the Dresselhaus SOI changes sign. Also notice that Eq. (4) implies that the anisotropic response is originated from the joint action of the Dresselhaus and Zeeman perturbations on the Rashba system.

Additionally, a phenomenological discussion on the role of disorder in the conductance and, particularly, resistance (better suited in experiments) can be found in Appendix A3.

## IV. EXPERIMENTS

The experimental setup consist of a top-gate-attached 40×40 ring array (ring radius $r$= 610 nm) fabricated by electron beam lithography and reactive ion etching. A scanning electron microscope image of the array is shown in Fig. 2(a). The ring array was covered with a 200-nm thick $Al_2O_3$ insulator made by atomic layer deposition and a Cr/Au top gate electrode in order to control the Rashba SOI strength alpha. All the measurements were performed at a temperature of 1.7 K.

We employed an InGaAs QW epitaxially grown on an InP (001) substrate. The detailed layer structure of the QW consists of, from the bottom, $In_{0.52}Al_{0.48}As$ (200 nm, buffer layer)/$In_{0.52}Al_{0.48}As$ (6 nm, carrier supply layer; Si-doping concentration of 4 x $10^{18}$ cm$^{-3}$) /$In_{0.52}Al_{0.48}As$ (15 nm, spacer layer) /$In_{0.53}Ga_{0.47}As$ (2.5 nm, QW)/$In_{0.73}Ga_{0.27}As$ (10 nm, QW) / $In_{0.53}Ga_{0.47}As$ (2.5 nm, QW)/ InP (5 nm, stopper layer) /$In_{0.52}Al_{0.48}As$ (20 nm, barrier layer)/AlAs (1.5 nm, barrier layer) /$In_{0.52}Al_{0.48}As$(5 nm, cap layer). The potential profiles of the QW are shown in Fig. 2(b). The electron wave function is almost confined in the $In_{0.53}Ga_{0.47}As$ and $In_{0.73}Ga_{0.27}As$ layers. By applying a negative gate voltage, the potential gradient is enhanced and the Rashba SOI is increased. The carrier density dependence of the Rashba SOI parameter is obtained from the analysis of the beating patterns of the Shubnikov- de Haas oscillations as a function of the gate voltage [Appendix B].

A common strategy is to investigate the gate-voltage dependence of the Al'tshuler-Aronov-Spivak (AAS) [17] oscillations amplitude, originated from the interference of time-reversal (TR) paths in the absence of magnetic flux (i.e., for vanishing perpendicular magnetic field $B_\perp = 0$). The phase contribution from the orbital part of the wave function to TR-path interference is always constructive at $B_\perp = 0$. Therefore, the AAS amplitude dependence on voltage reflects a phase contribution from the spin part of the wave function. This gives access to the AC spin-interference effect independently from the orbital phases at any gate-voltage value.

Ensemble averaging in the ring-array structure leads to clear AAS-interference patterns in transport measurements. We focused on AC spin interference under in-plane magnetic fields of variable strength $B_{//}$ and direction $\varphi$, defined with respect to the [100] axis. The magnetoresistance (MR) for fixed gate voltage and $B_{//}$= 1 T was measured for different orientations $\varphi$. Figure 3 (a) shows the results corresponding to a carrier density $N_S$= 1.9 x $10^{16}$ m$^{-2}$ and $\alpha$ = -1.5 x $10^{-12}$ eVm. The MR data includes AAS oscillations and background MR. The AAS amplitude shows a $\varphi$-angle dependence, with the maximum and minimum appearing at $\varphi = \pi/4$ and $\varphi = 3\pi/4$, respectively. For the sake of clarity, filtered AAS oscillations are presented in Fig. 3 (b). As expected from perturbation-theory and numerical analysis [27], the $\varphi$-angle dependence of the AAS amplitude has a π-periodicity. This angle dependence cannot be explained by the sole action of the Rashba SOI. Indeed, the observed anisotropy reflects the coexistence of Rashba and Dresselhaus SOIs. Corresponding sets of data at a different gate voltage with $N_S$= 1.52 x $10^{16}$ m$^{-2}$ ($\alpha$ = -2.8 x $10^{-12}$ eVm) are shown in Fig. 4 (a) and Fig. 4 (b). A weaker damping of the AAS oscillations as a function of the perpendicular magnetic field $B_\perp$ is explained by the narrowing of the effective channel width for decreasing carrier densities [28]. Most importantly, for $\alpha$ = -2.8 x $10^{-12}$ eVm (Fig. 4(b)) the AAS amplitude shows its minimum at $\varphi = \pi/4$ and its maximum at $\varphi = 3\pi/4$, a response opposite to the one observed at $\alpha$ = -1.5 x $10^{-12}$ eVm (Fig. 3(b)). This demonstrates the Rashba-SOI-induced anisotropy inversion without changing the sign of the Rashba SOI.

To study the observed inversion of the anisotropic response, in Fig. 5 (a) we show detailed experimental data on the Zeeman field angle dependence of the AAS amplitude for a field strength $B_{//}$=1 T at two different Rashba SOI strengths. We find that the angle-dependent pattern inverts as $\alpha$ changes from -1.5 x $10^{-12}$ eVm to -2.8 x $10^{-12}$ eVm while α's sign remains constant. This is well accounted by perturbation theory, Eq. (4), where the anisotropy inversion is attributed to the AC phase $\sqrt{1+\tilde{\alpha}^2} - 1$ in $A_G$, sharing geometric and dynamic phase contributions [20], [29]. The AAS amplitude dependence on the Zeeman field angle (for a given $\alpha$) has also a hybrid geometric/dynamic phase origin via $\phi_A$ [Appendix A2]. We notice that a purely geometric spin-phase tuning by the Zeeman field's strength is possible at magic angles $\varphi$ =0, π (where $\phi_A$ vanishes) through $\phi_B$ [22].

In order to account for realistic conditions in our models beyond the limitations of perturbation theory, we resort to 2D numerical simulations of disordered multi-mode rings. We use the Kwant code [30] with a disorder potential corresponding to a mean-free path of 1.8 μm, which is shorter that the ring circumference 3.8 μm. This disorder is crucial to develop dominating AAS interference paths [17]. The calculation details are described in [23]. We assume a ring radius of 610 nm and a ring channel including 5 modes, with carrier density $N_s$= 1.52 x $10^{16}$ m$^{-2}$. The in-plane Zeeman energy is set to $g\mu_B B_{//}$ =0.17 meV, with g= 3 and $B_{//}$= 1T, while $\beta$ is fixed to 0.3 x $10^{-12}$ eVm. These parameters are very similar to those of the InGaAs QW used in the present experiment. The results, depicted in Fig. 5 (b), show that the maximum and minimum AAS amplitudes appear around $\varphi = \pi/4$ and $\varphi = 3\pi/4$ for $\alpha$ = -1.5 x $10^{-12}$ eVm, while this anisotropy is reversed for $\alpha$ = -2.8 x $10^{-12}$ eVm. This is in quite good agreement with the experimental results shown in Fig. 5 (a).

In Fig. 6 we present the AAS amplitude measured as a function of the gate voltage corresponding to two different in-plane field angles $\varphi = \pi/4$ (red) and $\varphi = 3\pi/4$ (blue) and field strengths $B_{//}$= 1 T (Fig. 6 (left)) and $B_{//}$=2 T (Fig. 6 (right)) for $B_\perp$=0. The oscillatory response as a function of $\alpha$ is due to the AC effect induced by spin phases in TR-path interference. The observed period is well reproduced by perturbation theory, Eq. (2), once the gate voltage dependence of $\alpha$ is taken into account. We find that the AC oscillation amplitude decreases by increasing $B_{//}$ from 1 T to 2 T. This is explained by the spin-induced dephasing effect, as discussed in Ref. [27] and experimentally confirmed in [31].

V. DRESSELHAUS SPIN-ORBIT INVERSION

Figure 7 (a) shows the measured AC resistance difference between $\varphi = \pi/4$ and $\varphi = 3\pi/4$ (i.e., the resistance anisotropy), which displays an oscillatory behavior as a function of $\alpha$. The essential features of these oscillations are well captured by the 2D numerical simulations, Fig. 7 (b), except for an additional sign inversion observed in the experimental data in the region of weak Rashba SOI, around $\alpha$ = -1.2 ~ -1.7 x $10^{-12}$ eVm (corresponding to a carrier density $N_s$ = 2.0 ~ 1.8 x $10^{16}$ m$^{-2}$). In contrast, both the 2D Kwant simulations performed at *constant* Dresselhaus SOI strength and 1D models using AAS paths [23] predict the first sign reversal around $\alpha$ = -2.5 x $10^{-12}$ eVm.

This discrepancy is remarkable. The most plausible reason for such an additional anisotropy reversal is a sign change in the Dresselhaus SOI, as expected from the two-fold symmetry of the effective field of Eq. (1) and the perturbation theory in Eq. (4). By taking into account higher order and strain induced Dresselhaus effects, one notices that the sign of the resulting renormalized linear Dresselhaus SOI can be controlled by modifying the carrier density [32]. The Dresselhaus SOI Hamiltonian $H_D$ including an additional strain term $H_{Strain}$ is given by

$$H_D + H_{Strain} \approx \gamma\{\langle k_z^2\rangle - k_F^2/4 + \widetilde{D}(\varepsilon_{zz} - \varepsilon_{xx})/\gamma\}k_F(-\sigma_x \cos\varphi_0 + \sigma_y \sin\varphi_0). \quad (5)$$

The renormalized linear Dresselhaus SOI strength is expressed by $\beta' = \gamma\{\langle k_z^2\rangle - k_F^2/4 + \widetilde{D}(\varepsilon_{zz} - \varepsilon_{xx})/\gamma\}$, with $\widetilde{D}$ the deformation potential, and $\varepsilon_{zz}$ and $\varepsilon_{xx}$ the strain components [33]. The value of $\beta'$ is controlled electrically by the carrier density through $k_F = \sqrt{2\pi N_s}$. The confinement wave vector $\langle k_z^2\rangle = \int \Psi^*(z)(-\partial^2/\partial z^2)\Psi(z)dz$ in the InGaAs QW is estimated to be $1.32 \times 10^{16}\ m^{-2}$ by solving the Poisson-Schrödinger equation self-consistently. The tensile $\varepsilon_{zz}$ and compressive $\varepsilon_{xx} = \varepsilon_{yy}$ strain components for In$_{0.73}$Ga$_{0.27}$As/InP are calculated to be $\varepsilon_{zz} = 1.36\ \%$ and $\varepsilon_{xx} = -1.32\ \%$ by employing a reliable three dimensional nano-device simulator [34] often used for the design of semiconductor devices. The value of $\widetilde{D}/\hbar$ for our present sample is also expected to be similar to that of In$_x$Ga$_{1-x}$As (x= 5- 7 %) on GaAs substrate [33] since the Dresselhaus SOI is originated from a dipole electric field of bulk crystal. Trusted values of the deformation potential coefficient $\widetilde{D}/\hbar$ for strain induced Dresselhaus SOI in In$_x$Ga$_{1-x}$As (x= 5- 7 %) on GaAs substrate run from $0.5 \times 10^4\ m/s$ to $1.5 \times 10^4\ m/s$ [33]. Recent experimental and theoretical studies have shown that a secure value of the Dresselhaus parameter $\gamma$ in GaAs and InGaAs is close to $\gamma = 10 \times 10^{-30}\ eVm^3$ [35]. By taking $\widetilde{D}/\hbar = 1.0 \times 10^4\ m/s$ [33] and $\gamma = 10 \times 10^{-30}\ eVm^3$ [35], the renormalized linear Dresselhaus SOI $\beta'$ including the strain term is plotted as a function of the carrier density in Fig. 8. From Fig. 8, we obtain the critical carrier density $1.95 \times 10^{16}\ m^{-2}$ at which $\beta'$ changes its sign. It is difficult to explain our result without considering the strain term as shown by the red dashed line. It should be emphasized that the critical density is not changed if the ratio between $\gamma$ and $\widetilde{D}/\hbar$ is preserved. This critical carrier density is consistent with the one corresponding to the additional anisotropy reversal in Fig. 7 (a) and supports the conclusion of a sign change of $\beta'$ in our experiment by electric means.

**VI. CONCLUSIONS**

Our experiment demonstrates the anisotropic response of the AC interference effect in an electronic spin interferometer under in-plane Zeeman fields of different orientations with the support of theoretical and numerical models. We show that gate-controlled resistance measurements provide crucial information about the SO fields allowing to clarify the origin of the anisotropy in these setups, including a plausible control of the strength and sign of the renormalized linear Dresselhaus SOI. At the same time, we identify attainable ways to manipulate spin dynamic and geometric phases. These findings may contribute to guide future investigations towards understanding SOI in realistic materials relevant to quantum technologies.

**Acknowledgements:**

This work was supported by the Japan Society for the Promotion of Science through Grant-in-Aid for Specially Promoted Research No. H1505699, Grant-in-Aid for Scientific Research (C) No. 17K05510, Grant-in-Aid for Innovative Areas No. JP15K21717, and Core-to-Core Program, and by Projects No. FIS2014-53385-P and No. FIS2017-86478-P (MINECO/FEDER, Spain). D.F. acknowledges additional support from the Marie Sklodowska-Curie Grant Agreement No. 754340 (EU/H2020). The 2D simulations were calculated using the HOKUSAI system provided by Advanced Center for Computing and Communication (ACCC) at RIKEN. J.N. and F.N. are grateful to Makoto Kohda for valuable discussions.

**APPENDIX A: NONDEGENERATE PERTURBATION THEORY OF ANISOTROPIC SPIN INTERFERENCE**

**1. Rashba 1D ring under joint Dresselhaus and Zeeman perturbations.**

The Hamiltonian for spin carriers with effective mass *m\** confined in a Rashba 1D ring of radius *r* (parametrized by the azimuthal angle $\eta$) is given by [19]

$$\mathcal{H}_0 = -\frac{\hbar\omega_0}{2}\frac{\partial^2}{\partial\eta^2} - i\frac{\hbar\omega_R}{2}(\cos\eta\,\sigma_x + \sin\eta\,\sigma_y)\frac{\partial}{\partial\eta} - i\frac{\hbar\omega_R}{4}(\cos\eta\,\sigma_y - \sin\eta\,\sigma_x) \quad (S1)$$

with frequencies

$$\omega_0 = \frac{\hbar}{m^*r^2} \qquad \omega_R = \frac{2\alpha}{\hbar r} \quad . \quad (S2)$$

The main contributions to (S1) are the kinetic energy (first term) and the Rashba spin-orbit coupling (second term), corresponding to an effective (momentum-dependent) magnetic field pointing along the radial direction. The third term is the Meijer's correction [19] that guarantees the hermiticity of the Hamiltonian. The latter can be neglected in the semiclassic limit of large Fermi momentum, typically satisfied in mesoscopic semiconductors.

The unperturbed eigenstates $|n, \lambda, s\rangle_0$ (orbital quantum number *n*, travel direction $\lambda = \pm 1$, spin $s = \pm 1$) and eigenenergies $E_{\lambda n}^{0s}$ of $H_0$ are [20,22]

$$|n, +, \uparrow\rangle_0 = \exp(in\eta)\begin{pmatrix}\sin\theta/2 \\ e^{i\eta}\cos\theta/2\end{pmatrix} \qquad |n, +, \downarrow\rangle_0 = \exp(in\eta)\begin{pmatrix}\cos\theta/2 \\ -e^{i\eta}\sin\theta/2\end{pmatrix}, \quad (S3)$$

$$|n,-,\uparrow\rangle_0 = exp(-in\eta)\begin{pmatrix}\cos\theta/2\\-e^{i\eta}\sin\theta/2\end{pmatrix} \qquad |n,-,\downarrow\rangle_0 = exp(-in\eta)\begin{pmatrix}\sin\theta/2\\e^{i\eta}\cos\theta/2\end{pmatrix} \qquad (S4)$$

$$E_{\lambda n}^{0\,s} = \frac{\hbar\omega_0}{2}\left[\left(\lambda n + \frac{1}{2}\right)^2 + \frac{1}{4} + s\left|\lambda n + \frac{1}{2}\right|\sqrt{1+\tilde{\alpha}^2}\right], \qquad (S5)$$

with

$$\tilde{\alpha} = \frac{\omega_R}{\omega_0} = \frac{2\alpha m^* r}{\hbar^2} = \tan\theta \quad \text{and} \quad n \equiv kr \geq 0. \qquad (S6)$$

Notice that the unperturbed spin eigenstates (S3) and (S4) precess around the poles of the Bloch sphere by describing uniform cones with polar angle $\theta$ that subtend solid angles $\Omega_{\lambda n}^{0s} = 2\pi(1+\lambda s\cos\theta)$. We perturb the radial magnetic texture in $\mathcal{H}_0$ by introducing an in-plane Zeeman term $\Delta\mathcal{H}_1$ and a Dresselhaus spin-orbit term $\Delta\mathcal{H}_2$ of the form

$$\Delta\mathcal{H}_1 = \frac{\hbar\omega_B}{2}(\cos\varphi\,\sigma_x + \sin\varphi\,\sigma_y) \qquad (S7)$$

$$\Delta\mathcal{H}_2 = -i\frac{\hbar\omega_D}{2}(\sin\eta\,\sigma_x + \cos\eta\,\sigma_y)\frac{\partial}{\partial\varphi} - i\frac{\hbar\omega_D}{4}(\cos\eta\,\sigma_x - \sin\eta\,\sigma_y), \qquad (S8)$$

with $\omega_B = \frac{2\mu_B B_\parallel}{\hbar}$ and $\omega_D = \frac{2\beta}{\hbar r}$. $\qquad$ (S9)

The angle $\varphi$ in $\Delta\mathcal{H}_1$ defines the direction of the in-plane Zeeman field $B_\parallel$ with respect to the $x$ axis (coinciding with the crystallographic direction [100] in InGaAs heterostructures). The underlying anisotropy due to the coexistence of the Rashba and Dresselhaus SOIs is revealed as an explicit dependence on the angle $\varphi$ in the perturbed eigenenergies. The second term in $\Delta\mathcal{H}_2$ corresponds to a Meijer's-like correction to the Dresselhaus coupling.

By following the standard perturbation theory for nondegenerate systems [36] we find the first signs of anisotropy in the perturbed eigenenergies $E_{\lambda n}^S$ only after a 3rd-order expansion in $\Delta\mathcal{H} = \Delta\mathcal{H}_1 + \Delta\mathcal{H}_2$. This procedure leads to

$$E_{\lambda n}^S = E_{\lambda n}^{0\,S} + s\frac{(\hbar\omega_B)^2}{\hbar\omega_0}\frac{1}{8n|\tilde{\alpha}|} + s\frac{(\hbar\omega_D)^2}{\hbar\omega_0}\frac{n}{8|\tilde{\alpha}|} - s\frac{(\hbar\omega_B)^2\hbar\omega_D}{(\hbar\omega_0)^2}\frac{\text{sign}[\tilde{\alpha}]}{16\,n\,\tilde{\alpha}^2}\sin(2\varphi). \qquad (S10)$$

The anisotropic response of the perturbed eigenenergies (S10) to the Zeeman field orientation $\varphi$ appears at the 1st order in the Dresselhaus coupling strength and at the 2nd order in Zeeman one, showing that the anisotropy can discriminate the sign of the Dresselhaus term. The perturbed eigenstates $|n, \lambda, s\rangle$ need to be expanded only up to 2$^{nd}$ order in $\Delta\mathcal{H} = \Delta\mathcal{H}_1 + \Delta\mathcal{H}_2$ to show the first anisotropic features due to the joint Dresselhaus-Zeeman action. Up to a normalization factor, they read

$$|n, \lambda, s\rangle = Z_{\lambda n}^S |n, \lambda, s\rangle_0 \Bigg[ \Bigg( 1 - \sum_{q=-2}^{2}\sum_{s'=\pm 1} \frac{{}_0\langle n,\lambda,s|\Delta H|n+q,\lambda,s'\rangle_0\ {}_0\langle n+q,\lambda,s'|\Delta H|n,\lambda,s\rangle_0}{2\left(E_{\lambda n}^{0s} - E_{\lambda(n+q)}^{0s'}\right)^2} \Bigg) +$$

$$\frac{\omega_B}{4\omega_0 n}\Bigg(\lambda s e^{i\lambda\varphi}|n-1,\lambda,s\rangle_0 - \frac{e^{i\lambda\varphi}}{|\tilde{\alpha}|}|n-1,\lambda,\bar{s}\rangle_0 - \lambda s e^{-i\lambda\varphi}|n+1,\lambda,s\rangle_0 + \frac{e^{-i\lambda\varphi}}{|\tilde{\alpha}|}|n-1,\lambda,\bar{s}\rangle_0\Bigg) +$$

$$i\frac{\omega_D}{4\omega_0}\Bigg(\frac{\lambda s}{2}|n-2,\lambda,s\rangle_0 - \frac{1}{|\tilde{\alpha}|}|n-2,\lambda,\bar{s}\rangle_0 + \frac{\lambda s}{2}|n+2,\lambda,s\rangle_0 - \frac{1}{|\tilde{\alpha}|}|n+2,\lambda,\bar{s}\rangle_0\Bigg) +$$

$$\sum_{p=-4}^{4}\sum_{s'=\pm 1}|n+p,\lambda,s'\rangle_0 \sum_{q=-2}^{2}\sum_{s''=\pm 1} \frac{{}_0\langle n+p,\lambda,s'|\Delta H|n+q,\lambda,s''\rangle_0\ {}_0\langle n+q,\lambda,s''|\Delta H|n,\lambda,s\rangle_0}{\left(E_{\lambda n}^{0s} - E_{\lambda(n+p)}^{0s'}\right)\left(E_{\lambda n}^{0s} - E_{\lambda(n+q)}^{0s''}\right)} \Bigg],$$

(S11)

where the sums in (S11) run such that the denominators do not vanish. The perturbative corrections to the first term in (S11) lead to $\omega_B{}^2$ and $\omega_D{}^2$ contributions, only, while the last term shows additional $\omega_B\omega_D$ joint contributions. We point out that the results (S10) and (S11) hold for $1 \ll \tilde{\alpha} \ll 2n$, where degeneracy mixing is avoided and the perturbative approach is sound.

**2. Anisotropic conductance and the role of geometric/dynamic spin phases.**

We calculate the AAS (Al'tshuler-Aronov-Spivak) corrections to the conductance of a two-terminal AC 1D ring originated from the interference of time-reversed paths at the lowest order (i.e., semiclassical paths describing single windings around the ring corresponding to strongly coupled contacts) by following a procedure similar to our previous works on Rashba rings [20, 22], where the phase difference gathered by counter-propagating spin carriers in found by solving $E_{\lambda n}^S = E_F$ for non-integer orbital numbers $n_\lambda^s$, with $E_F$ the Fermi energy. As a result, the AAS conductance takes the general form $G_{AAS} = \left(\frac{e^2}{h}\right)\left(1 + [\cos 2\pi(n_-^\downarrow - n_+^\uparrow) + \cos 2\pi(n_-^\uparrow - n_+^\downarrow)]/2\right)$ with $n_-^\downarrow - n_+^\uparrow = 1 +$ $\sqrt{1+\tilde{\alpha}^2} + \phi_B + \phi_D + \phi_A$ and $n_-^\uparrow - n_+^\downarrow = 1 - \sqrt{1+\tilde{\alpha}^2} - \phi_B - \phi_D - \phi_A$. We then find

$$G_{AAS} = \frac{e^2}{h}\left[1 - \cos\left\{2\pi\left(\sqrt{1+\tilde{\alpha}^2} - 1 + \phi_B + \phi_D + \phi_A\right)\right\}\right] \quad (S12)$$

with

$$\phi_B = \left(\frac{\omega_B}{\omega_0}\right)^2 \frac{1}{4(k_F r)^2 |\tilde{\alpha}|} \quad (S13)$$

$$\phi_D = \left(\frac{\omega_D}{\omega_0}\right)^2 \frac{1}{4|\tilde{\alpha}|} \quad (S14)$$

$$\phi_A = -\left(\frac{\omega_B}{\omega_0}\right)^2 \left(\frac{\omega_D}{\omega_0}\right) \frac{\text{sign}[\tilde{\alpha}]}{8(k_F r)^2 \tilde{\alpha}^2} \sin 2\varphi \quad (S15)$$

These contributions represent a Zeeman phase shift $\phi_B$, a Dresselhaus phase shift $\phi_D$, and an anisotropic phase shift $\phi_A$. The latter depends explicitly on $\varphi$, showing the two-fold symmetry anticipated in Eq. (1) with opposite extreme values at $\pi/4$ and $3\pi/4$. Notice that $\phi_B$ and $\phi_D$ derive from the quadratic contributions to the perturbed eigenenergies (S10). Hence, according to perturbation theory [36, 37], they originate from the linear contributions to the perturbed eigenstates (S11). As for $\phi_A$, it is a consequence of the cubic contributions to (S10) and the quadratic one to (S11). Each of the phases (S13-S15) can be of either pure or hybrid geometric/dynamic origin. The geometric-phase contribution $\phi_g$ to the conductance (S12) can be evaluated from the perturbed eigenstates (S11) as [22, 25]

$\phi_g = \frac{i}{\pi}\int_0^{2\pi} d\eta \langle \overline{n,\lambda,s} | \partial_\eta | \overline{n,\lambda,s} \rangle$

$= -(1+\lambda s \cos\theta) - \lambda s \phi_B - \frac{\lambda s}{2}\phi_D - \frac{\lambda s}{2}\phi_A \quad (S16)$

$= -\frac{1}{2\pi}\left(\Omega_{\lambda n}^{0s} + \Delta\Omega_{\lambda n}^s\right)$

with $|\overline{n,\lambda,s}\rangle \equiv exp(-i\lambda n\eta)|n,\lambda,s\rangle$. The first term in (S16), $-(1+\lambda s \cos\theta) = -\Omega_{\lambda n}^{0s}/2\pi$, can be easily identified with the geometric phase contribution to the unperturbed AC phase $\sqrt{1+\tilde{\alpha}^2} - 1$ in (S12) by choosing $\lambda = 1$ and $s = -1$. The complementary dynamic-phase contribution to the unperturbed AC phase reads $\tilde{\alpha}\sin\theta$, identified with the spin eigenstate projection on the local in-plane field [20,22]. The additional contributions to the geometric phase in (S16) are interpreted as perturbations $\Delta\Omega_{\lambda n}^s$ to $\Omega_{\lambda n}^{0s}$. The share of these geometric-phase contributions in the conductance (S12) depend on the corresponding weight factors appearing in (S16). The Zeeman contribution $\phi_B$ to (S12) results to be of purely geometric origin (as reported in [22]) as a consequence of a weight factor 1 (absolute value) in (S16). The Dresselhaus contribution $\phi_D$ to (S12), with a weight factor ½ in (S16), turns out to be only 50% geometric (the other 50% is of dynamic origin), likely due to the different symmetry class of Zeeman and Dresselhaus perturbations. As for the geometric-phase contribution to the anisotropic phase $\phi_A$, the corresponding weight factor ½ in (S16) indicates a 50% share. Namely, $\phi_A$ has an hybrid geometric/dynamical origin.

### 3. Role of disorder.

The role of disorder can be effectively accounted by introducing a classical conductance $G_0$ and a quantum-correction amplitude $a \ll 1$ such that

$$G_{AAS} \approx G_0[1 - a\cos\{2\pi(\sqrt{1+\tilde{\alpha}^2} - 1 + \phi_B + \phi_D + \phi_A)\}]. \tag{S17}$$

The resistance $R_{AAS} = 1/G_{AAS}$, better suited in experiments, then reads

$$R_{AAS} \approx R_0[1 + a\cos\{2\pi(\sqrt{1+\tilde{\alpha}^2} - 1 + \phi_B + \phi_D + \phi_A)\}] \tag{S18}$$

with $R_0 = 1/G_0$ the classical resistance.

By noticing that the anisotropic phase $\phi_A$ is much smaller than the unperturbed AC phase $\sqrt{1+\tilde{\alpha}^2} - 1$, we rewrite the resistance as

$$\frac{R_{AAS}}{R_0} \approx A_1 + A_2 \sin(2\varphi) \tag{S19}$$

with
$$A_1 = 1 + a\cos\{2\pi(\sqrt{1+\tilde{\alpha}^2} - 1 + \phi_B + \phi_D)\}, \tag{S20}$$

$$A_2 = a\, 2\pi \left(\frac{\omega_B}{\omega_0}\right)^2 \left(\frac{\omega_D}{\omega_0}\right) \frac{\text{sign}[\tilde{\alpha}]}{8(k_F r)^2 \tilde{\alpha}^2} \sin\{2\pi(\sqrt{1+\tilde{\alpha}^2} - 1 + \phi_B + \phi_D)\}. \tag{S21}$$

The Eq. (S19) shows an anisotropic response of the resistance to the Zeeman field's direction $\varphi$. Moreover, the sign of the anisotropy can be independently modulated by the Rashba strength $\tilde{\alpha}$ but its response is isotropic.

### APPENDIX B: CARRIER DENSITY DEPENDENCE OF RASHBA SOI STRENGTH

The gate fitted Hall bar (70 μm x 280 μm) was fabricated on the same chip on which the spin interferometer (40 x 40 ring array) was put. The relation between carrier density and Rashba SOI strength was obtained from the analysis of Shubnikov-de Haas (SdH) oscillations as shown in Fig. 9. The SdH oscillations show a beating pattern because of spin splitting due to the strong Rashba SOI. The Rashba SOI strength is given by

$$\alpha = \frac{\hbar^2(\sqrt{2\pi N_\uparrow}-\sqrt{2\pi N_\downarrow})}{2m^*} \quad (S22)$$

Here, $N_\uparrow$ and $N_\downarrow$ are the spin split densities, which can be obtained from the fast Fourier Transform (FFT) spectra of SdH oscillations. The electron effective mass $m^* = 0.05$ can be estimated by analyzing the temperature dependence of SdH oscillation amplitude. The relation between the Rashba SOI parameter $\alpha$ and the carrier density $N_s$ is plotted in Fig. 10. In the above analysis, we assumed that the Dresselhaus SOI strength is negligible since $\beta_D = \gamma\langle k_z^2\rangle$ is one order of magnitude smaller than the Rashba SOI strength.

Fig. 1

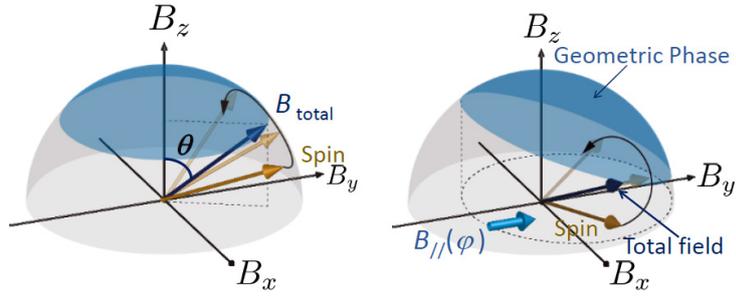

Fig. 1 Schematic illustration of spin geometric and dynamic phases in an AC spin interferometer. (Left) In the moving electron's rest frame, the SOI field subtends a solid angle (blue) in a round trip around the interference ring. The solid angle is proportional to the spin geometric phase. Only when the Lamor frequency of spin precession $\omega_s$ is fast enough compared with the orbital frequency $\omega_c$, the SOI field is in x-y plane (adiabatic limit). Spin precession around SOI field $B_{\text{total}}$ is associated with the dynamical phase. The angle $\theta$ is given by the relation $\tan\theta = \omega_s/\omega_c$. (Right) The in-plane field modulates the geometric phase by changing the solid angle subtended by the total effective field.

Fig. 2 (a) 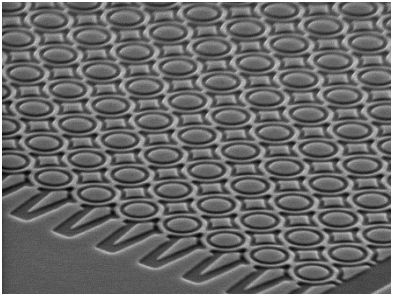 Fig. 2 (b) 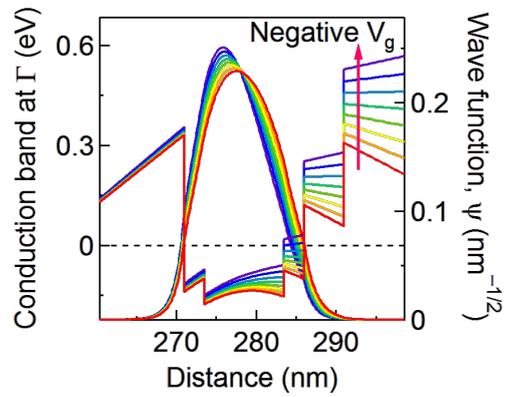

Fig. 2 (a) Scanning electron microscope image of an array of 40 x 40 InGaAs-based rings. The radius of each ring is 610 nm. (b) Calculated potential energies relative to the Fermi energy (left scale) and the squared wave functions (right scale) for the samples used in this paper. The wave functions are confined in the InGaAs quantum well. The potential gradient becomes lager with increasing negative gate voltage, resulting in an enhancement of the Rashba SOI strength.

Fig. 3 (a) 

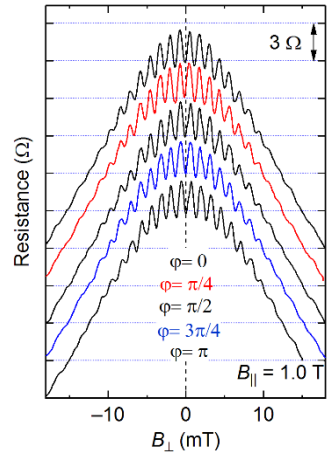 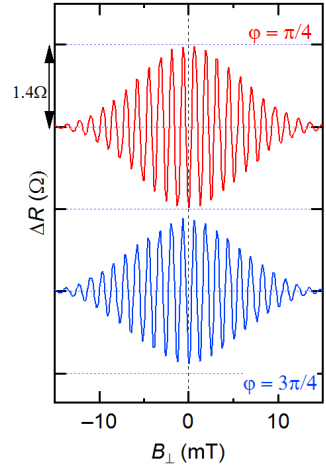

Fig. 3 (a) AAS oscillations with fixed Rashba SOI strength $\alpha = -1.5 \times 10^{-12}$ eVm measured by varying the in-plane magnetic field direction for a constant field strength $B_{//} = 1$ T. (b) Filtered AAS oscillations. The amplitude of the AAS oscillations shows an angle dependence, with the maximum and minimum at $\varphi = \pi/4$ and $\varphi = 3\pi/4$, respectively.

Fig. 4 (a)    Fig. 4 (b)

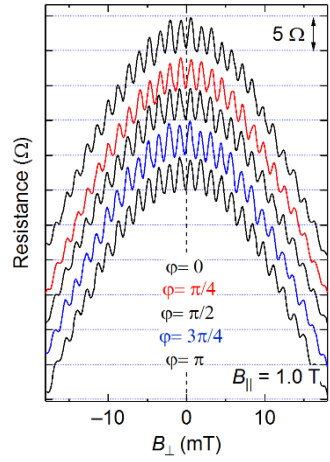 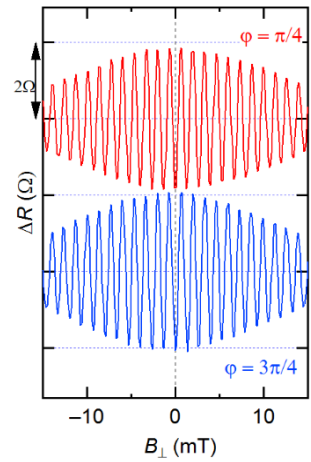

Fig. 4 (a) AAS oscillations corresponding to a fixed Rashba SOI strength $\alpha = -2.8 \times 10^{-12}$ eVm measured by varying the in-plane magnetic field direction for a constant field strength $B_{//} = 1$ T. (b) Filtered AAS oscillations. The extrema of the AAS oscillation amplitude are switched when compared with the case shown in Fig. 3(a), with the maximum at $\varphi = 3\pi/4$ and the minimum at $\varphi = \pi/4$.

Fig. 5 (a) <span style="float:right">Fig. 5 (b)</span>

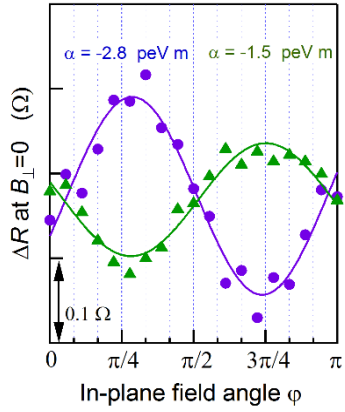 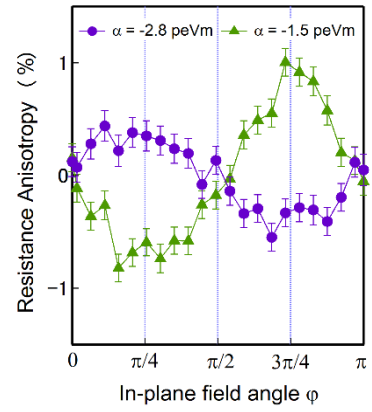

Fig. 5 (a) Angle dependence of the AAS amplitudes at $B_\perp = 0$ in the presence of an in-plane field $B_{//}= 1$ T. For a Rashba SOI strength $\alpha \approx -1.5 \times 10^{-12}$ eVm, the maximum and minimum appear around $\varphi = \pi/4$ and $\varphi = 3\pi/4$, respectively. The anisotropic response inverts for $\alpha \approx -2.8 \times 10^{-12}$ eVm. (b) Corresponding 2D numerical simulations with realistic parameters. The results are in good agreement with the experimental data.

Fig. 6

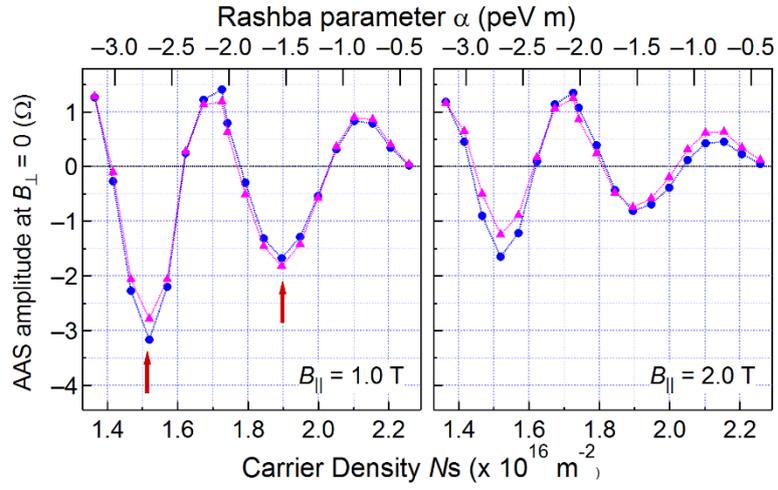

Fig. 6 Gate-voltage dependence of the AAS oscillation amplitude at $B_\perp = 0$ under in-plane fields $B_{//}= 1$ T (left) and $B_{//}= 2$ T (right) applied along different directions. The AAS amplitude modulation by the Rashba SOI strength arises from the AC spin interference. An anisotropic response is observed at in-plane field angles $\varphi = \pi/4$ (triangle) and $\varphi = 3\pi/4$ (circle). For an in-plane field $B_{//}= 1$ T, the anisotropy is reversed by tuning the Rashba SOI strength $\alpha$ while keeping its sign unchanged, as shown by arrows.

Fig. 7 (a) 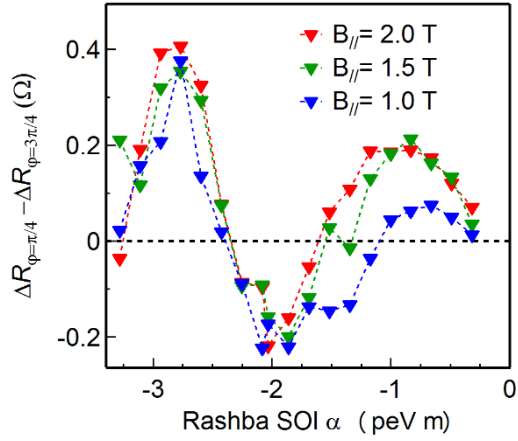 Fig. 7 (b) 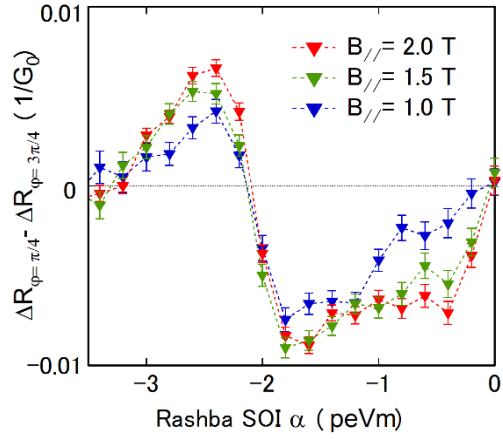

Fig. 7 (a) AC resistance difference (anisotropy) between in-plane field orientations $\varphi = \pi/4$ and $\varphi = 3\pi/4$ as a function of the Rashba SOI parameter $\alpha$. The anisotropy shows an oscillatory behavior as a function of $\alpha$. (b) Corresponding 2D numerical simulations. A remarkable discrepancy appears around $\alpha = -1.2 \sim -1.7 \times 10^{-12}$ eVm, interpreted as a (gate-controlled) sign change of the renormalized linear Dresselhaus SOI (see text).

Fig. 8

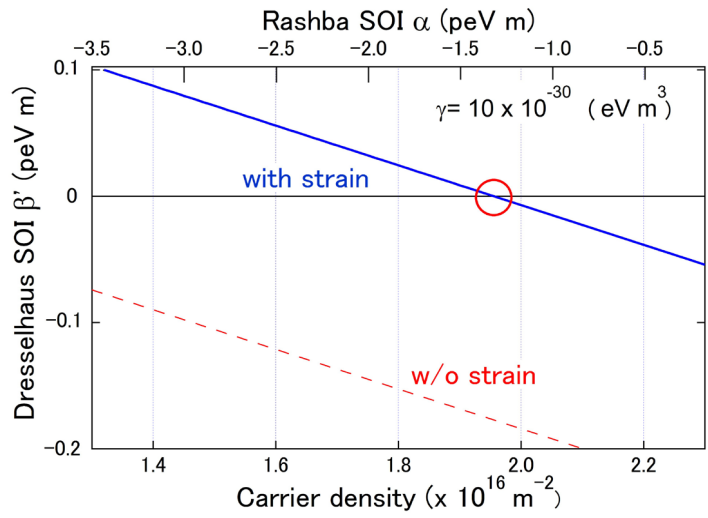

Fig. 8 Renormalized linear Dresselhaus SOI $\beta'$ including the strain induced term as a function of the carrier density. Red dashed line does not include the strain term.

Fig. 9

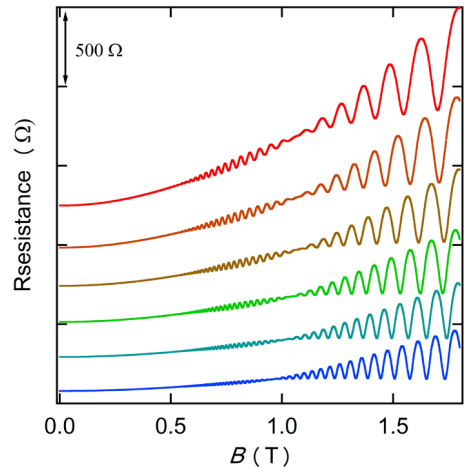

Fig. 9 Gate voltage dependence of the SdH oscillations. From top to bottom, the corresponding carrier densities are 0.8, 1.0, 1.2, 1.35, 1.55, and 1.7 x $10^{16}$ m$^{-2}$.

Fig. 10

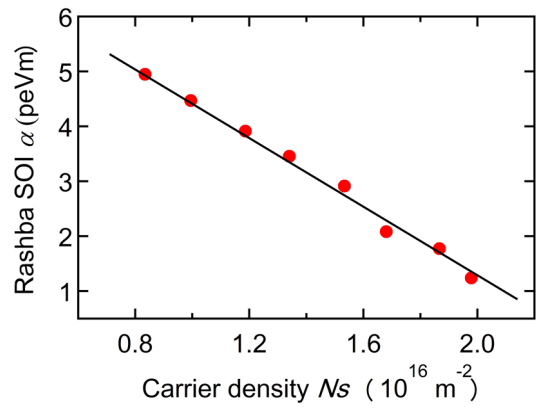

Fig. 10 Relation between Rashba SOI parameter and carrier density.